\documentclass{emulateapj}
\usepackage{verbatim}
\usepackage[breaklinks,colorlinks,citecolor=blue,urlcolor=blue]{hyperref}
\usepackage[all]{hypcap}
\usepackage{float}
\newcommand{\angstrom}{\mbox{\normalfont\AA}}
\include{package}

\shorttitle{TEMPORAL VARIABILITY TOWARD THE MONOCEROS LOOP}
\shortauthors{DIRKS \& MEYER}

\begin{document}

\title{Temporal Variability Of Interstellar N\lowercase{a~\textsc{i}} Absorption Toward The Monoceros Loop}

\author{Cody Dirks\altaffilmark{1} and David M. Meyer\altaffilmark{1}}
\affil{Center for Interdisciplinary Research and Exploration in Astrophysics, Department of Physics \& Astronomy,\\ Northwestern University, 2145 Sheridan Road, Evanston, IL 60208, USA; \\ codydirks2017@u.northwestern.edu, davemeyer@northwestern.edu}

\altaffiltext{1}{Visiting Astronomer, Kitt Peak National Observatory, National Optical Astronomy Observatory, which is operated by the Association of Universities for Research in Astronomy, Inc., under cooperative agreement with the National Science Foundation.}

\begin{abstract}
We report the first evidence of temporal variability in the interstellar Na\,\,\textsc{i} absorption toward HD~47240, which lies behind the Monoceros Loop supernova remnant (SNR). Analysis of multi-epoch Kitt Peak coud\'{e} feed spectra from this sightline taken over an eight-year period reveals significant variation in both the observed column density and the central velocities of the high-velocity gas components in these spectra. Given the $\sim$1.3\,\,mas\,\,yr$^{-1}$ proper motion of HD~47240 and an SNR distance of 1.6 kpc, this variation would imply $\sim$10\,\,au fluctuations within the SNR shell. Similar variations have been previously reported in the Vela SNR, suggesting a connection between the expanding SNR gas and the observed variations. We speculate on the potential nature of the observed variations toward HD~47240 in the context of the expanding remnant gas interacting with the ambient interstellar medium.
\end{abstract}


\section{Introduction}

\begin{figure*}[!tp]
\centering
\includegraphics[scale=0.4]{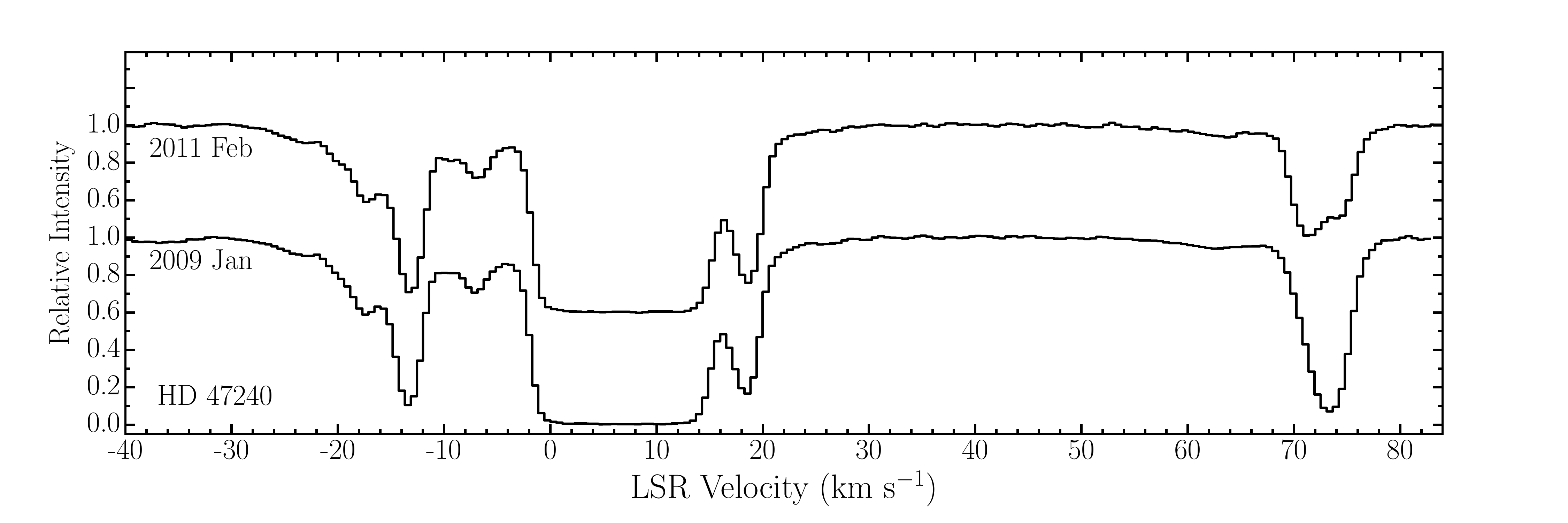}
\vspace{-15pt}
\caption{Normalized multi-epoch spectra of Na\,\textsc{i} D2 lines at $\lambda\;5890$ toward HD~47240 taken at \emph{R} $\sim$230,000 ($\Delta v \sim$1.3\,km\,s$^{-1}$). The multiple absorption components between $-30$ km s$^{-1}$ and $+30$ km s$^{-1}$ do not exhibit any variation across our range of observations, and thus are considered constant and not included in our analysis of temporal variation, which is clearly seen in the HV components near \mbox{$+70$ km\,s$^{-1}$}.}
\label{fig:full}
\end{figure*}

The diffuse interstellar medium (ISM) contains within it a rich variety of structure, ranging from long, slender filaments \citep{clark14}, to sheet-like structures \citep{heiles03}, to the well-known shells and supershells surrounding stellar associations. Still unknown, however, is the extent of small-scale structure within these common, larger entities. Recently, the high pressures found within the Local Leo Cold Cloud \citep{peek11,meyer12} has shown that structure on the scale of $\sim$200\,au exists in the cold, diffuse H\,\textsc{i} within our own Local Bubble. The mechanism by which these small-scale features form is still unknown, but may be similar to the formation of filamentary structures via turbulent gas flows within molecular clouds \citep{andre14}. Based on temporal variation of absorption in H\,\textsc{i} 21\,\,cm emission from background quasars, \citet{heiles97} found these small-scale structures to be significantly over-pressured with respect to their surroundings if a spherical geometry was assumed, a result in stark contrast with the long-standing view of an ISM in pressure equilibrium \citep{mckee77}. Heiles rectified this problem by invoking nonspherical geometries (such as cylinders and disks, or more complicated filaments and sheets) for these small-scale structures. Alternatively, \citet{deshpande00} argued that observations of small-scale structure in H\,\textsc{i} are a natural consequence of hierarchical structure in the ISM. 

Further observations in the optical have found a variety of sightlines exhibiting spatial or temporal variations \citep{meyer96,crawford00,price00,lauroesch03,points04}, primarily in the Na\,\textsc{i} D lines. However, \citet{lauroesch03} highlight a particular issue with using Na\,\textsc{i} D lines to detect variation in the ISM. Since Na\,\textsc{i} is not the dominant ion of Na in the ISM, the observed variations in \emph{N}(Na\,\textsc{i}) may not be reflective of changes in the total H\,\textsc{i} column, but rather the result of changes in the physical conditions of the absorbing gas, which leads to an enhancement or dehancement in the Na\,\textsc{i} trace neutral. Thus, it remains unclear if the many variations seen in the optical are indicative of corresponding variations in the H\,\textsc{i} gas structure, or, rather, of changes in the physical properties of the gas.

Supernova remnants (SNRs) have long been an interesting laboratory for studying the ISM. Recently, \citet{kameswara15} reported on the variability of a number of ISM absorption components toward several stars in the Vela SNR, suggesting a connection between ISM variability and SNRs. Here we focus on the Monoceros Loop SNR, a well-evolved SNR ($\sim$100,000 years old) with an expansion velocity of $\sim$50\,km\,s$^{-1}$, derived from absorption seen in the approaching and receding edges of the shell at \mbox{$-31$\,km\,s$^{-1}$} and \mbox{$+69$\,km\,s$^{-1}$}, from the stars HD~47359 and HD~47240, respectively \citep{wallerstein76}. Furthermore, \citet{odegard86} determined that the SNR is within the Mon OB2 association, and also interacting with the Rosette Nebula, placing it at a distance of 1.6 kpc. At this distance, the SNR would have a linear diameter of roughly 110\,pc. 

Analysis of the HD~47240 sightline at 12\,km\,s$^{-1}$ spectral resolution by \citet{wallerstein76} found atypically low N(Na\,\textsc{i})/N(Ca\,\textsc{ii}) ratios for the high-velocity (HV) gas. This abnormal ratio was taken as evidence of dust grains being shocked and returning Ca\,\textsc{ii} (typically heavily depleted into grains) to the gas phase \citep{routly52}. More recent work by \citet{welsh01} utilized \emph{FUSE} spectra at a similar resolution, and detected absorption from many species of various ionization levels within the HV gas. Absorption was seen at $V_{LSR} \sim +65$\,km\,s$^{-1}$ for several low-ionization atomic states (O\,\textsc{i}, Ar\,\textsc{i}, N\,\textsc{i}, C\,\textsc{i}, Fe\,\textsc{ii}, and P\,\textsc{ii} in \emph{FUSE} spectra, and O\,\textsc{i}, Al\,\textsc{ii}, Si\,\textsc{ii}, S\,\textsc{ii}, Fe\,\textsc{ii}, C\,\textsc{ii}$^{*}$, Mg\,\textsc{i}, and Mg\,\textsc{ii} in \emph{IUE} spectra). They also noted the absence of a corresponding HV component in high-ionization states (O\,\textsc{vi} in their \emph{FUSE} spectrum, Si\,\textsc{iv}, C\,\textsc{iv} and Al\,\textsc{iii} in \emph{IUE} spectra).
They argued that the observed combination of ionized and neutral species could not physically coexist in the same cloud, and thus conjectured that the observed HV component is actually ``several ionized and neutral gas shells expanding at slightly different velocities away from the center of the SNR."
These same \emph{FUSE} spectra also reveal rotationally excited (\emph{J}=3) H$_2$ at a similar high velocity \citep{welsh02}, which can be produced via interstellar shocks.

In this paper, we present new multi-epoch high-resolution (\emph{R} $\sim$ 230,000) observations of HD~47240 taken with the Kitt Peak National Observatory  coud\'{e} feed telescope over an eight-year period. These data provide strong evidence that the high-velocity component associated with the Monoceros Loop is actually two distinct components, and that both components undergo dramatic variation in column density and central velocity over periods of a few years. The scale of these variations is determined by the combination of the proper motion of both the background star as well as the intervening gas. Based on these factors, we estimate the characteristic scale of variation to be a few tens of au.

\section{Observations and Data Reductions}
Observations of the interstellar Na\,\textsc{i} D2 $\lambda$5889.951 and D1 $\lambda$5895.924 absorption lines were obtained toward HD~47240 (a B1Ib star located at a distance of 1.8\,kpc) on five separate occasions over an approximately eight-year period (2006 January, 2008 March, 2009 January, 2011 February, 2014 January) . These data were gathered using the 0.9m coud\'{e} feed telescope and spectrograph at Kitt Peak National Observatory. The spectrograph was configured using the long focal length camera (camera 6), the echelle grating, a cross-dispersing grism (770), and the T2KB Tek2048 CCD detector. Using a 150\,$\mu$m slit, we achieved a 2.1 pixel spectral resolution of 0.026\,\angstrom, or 1.3\,km\,s$^{-1}$ at the Na\,\textsc{i} wavelengths, as measured with a Th--Ar calibration lamp. We then utilized the NOAO IRAF echelle data reduction package to correct our CCD exposures for scattered light and bias, and to flat-field our exposures. From these exposures, we then used the appropriate packages to extract spectral orders and wavelength calibrate our spectra.
We also applied these techniques to a nearby early-type star (\mbox{$\alpha$ Leo}) whose sightline is essentially devoid of any interstellar Na\,\textsc{i} absorption, allowing us to create a nightly template of telluric lines, which was used to divide out the telluric absorption in all HD~47240 spectra taken on the same night as the template.
Finally, we used the continuum fitting routines in the IRAF package to identify the stellar continuum in the vicinity of the Na\,\textsc{i} D lines, and normalized relative to this stellar continuum, leaving residual intensities due to interstellar absorption. All final spectra have S/N $>$ 100.

\capstartfalse
\begin{deluxetable}{crcr}[]
\tabletypesize{\scriptsize}
\tablecaption{Fit Parameters for Static Low-velocity N$\mathrm{a}$\,\textsc{i} Components toward HD~47240}
\tablehead{\colhead{Component} & \colhead{\emph{N} ($\times 10^{11}$\,cm$^{-2}$)}& \colhead{\emph{b} (km\,s$^{-1}$)}& \colhead{\emph{v} (km\,s$^{-1}$)}}
\startdata
1 & 0.52 $\pm$ 0.03 & 3.53 $\pm$ 0.24 & $-$22.00 $\pm$ 0.25\\
2 & 2.11 $\pm$ 0.03 & 2.41 $\pm$ 0.05 & $-$16.78 $\pm$ 0.05\\
3 & 6.15 $\pm$ 0.08 & 0.82 $\pm$ 0.01 & $-$12.96 $\pm$ 0.01\\
4 & 1.41 $\pm$ 0.02 & 2.67 $\pm$ 0.07 &  $-$7.39 $\pm$ 0.05\\
5&160.76 $\pm$ 25.11& 3.56 $\pm$ 0.01 &    +7.05 $\pm$ 0.01\\
6 & 1.14 $\pm$ 0.31 & 8.38 $\pm$ 0.58 &   +15.71 $\pm$ 1.31\\
7 & 4.61 $\pm$ 0.08 & 1.09 $\pm$ 0.02 &   +18.64 $\pm$ 0.01
\enddata
\tablecomments{Errors were obtained by fixing two of the three parameters and allowing the third to vary. Due to the large number ($\sim 40$) of simultaneous lines being fit, allowing all three parameters to vary simultaneously caused instabilities in the fitting routines.}
\end{deluxetable}
\capstarttrue

\begin{figure*}[!tp]
\centering
\includegraphics[scale=0.4]{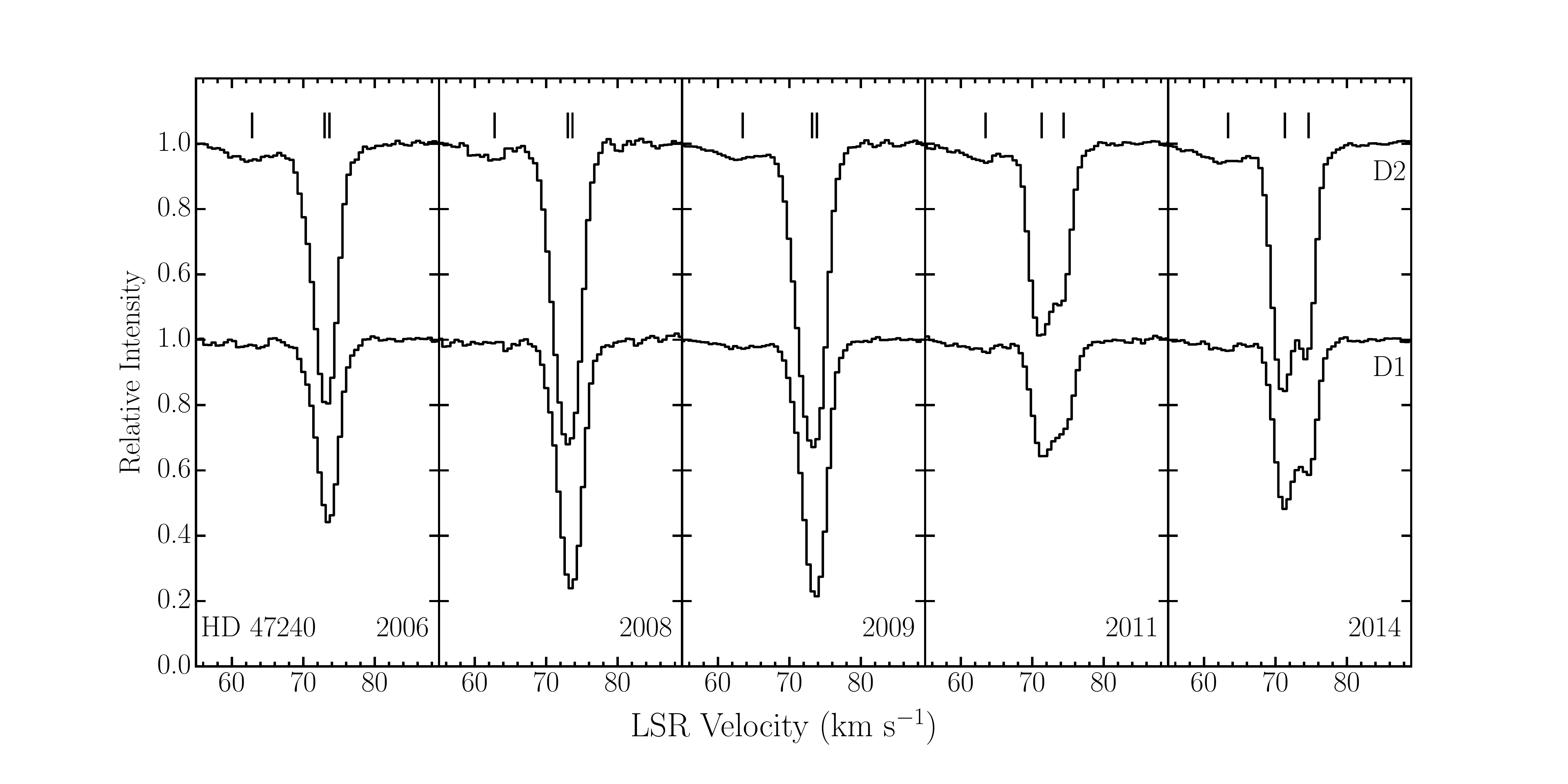}
\vspace{-20pt}
\caption{Isolated high-velocity Na\,\textsc{i} components toward HD~47240 for each of the five epochs in our sample, with the top row representing the D2 lines ($\lambda\;5889.951$) and the bottom row representing D1 ($\lambda\;5985.924$). Central velocities of the weak, broad component near \mbox{$+63$\,km\,s$^{-1}$} and the two components near \mbox{$+73$\,km\,s$^{-1}$} are marked by dashes above the spectra. Of particular note is the substantial increase in line depth between 2006 and 2008, and the substantial decrease in depth between 2009 and 2011. This decrease in absorption is further accompanied by the splitting of the line into two distinct velocity components, with this velocity separation continuing to increase in the 2014 spectrum.}
\label{fig:hv}
\end{figure*}

\capstartfalse
\begin{deluxetable*}{ccccrrrrr}[]
\centering
\tabletypesize{\scriptsize}
\tablecaption{Fit Parameters for High-velocity N$\mathrm{a}$\,\textsc{i} Components toward HD~47240}
\tablehead{\colhead{Parameter} & \colhead{Unit} & \colhead{Component} & \colhead{2006}& \colhead{2008}& \colhead{2009}& \colhead{2011}& \colhead{2014}}
\startdata

\emph{N} &$\times 10^{11}$\,cm$^{-2}$
  & Comp. 1 & 0.37 $\pm$ 0.08 & 0.30 $\pm$ 0.09 & 0.41 $\pm$ 0.06 & 0.45 $\pm$ 0.06 & 0.50 $\pm$ 0.07\\
& & Comp. 2 & 2.74 $\pm$ 0.10 & 4.99 $\pm$ 0.29 & 5.56 $\pm$ 0.15 & 2.64 $\pm$ 0.05 & 4.38 $\pm$ 0.03\\
& & Comp. 3 & 3.01 $\pm$ 0.10 & 4.91 $\pm$ 0.29 & 5.23 $\pm$ 0.17 & 2.03 $\pm$ 0.04 & 2.98 $\pm$ 0.03\\
& & & & & & \\

\emph{b} &km\,s$^{-1}$ 
  & Comp. 1 & 4.14 $\pm$ 0.16 & 4.17 $\pm$ 0.37 & 4.72 $\pm$ 0.17 & 4.78 $\pm$ 0.16 & 4.95 $\pm$ 0.14\\
& & Comp. 2 & 2.60 $\pm$ 0.04 & 2.24 $\pm$ 0.04 & 2.30 $\pm$ 0.02 & 1.39 $\pm$ 0.03 & 1.36 $\pm$ 0.01\\
& & Comp. 3 & 1.02 $\pm$ 0.03 & 1.07 $\pm$ 0.05 & 1.00 $\pm$ 0.03 & 1.45 $\pm$ 0.03 & 1.21 $\pm$ 0.02\\
& & & & & & \\

\emph{v} &km\,s$^{-1}$ 
  & Comp. 1 & 62.84 $\pm$ 0.11 & 62.79 $\pm$ 0.25 & 63.50 $\pm$ 0.11 & 63.48 $\pm$ 0.10 & 63.37 $\pm$ 0.09\\
& & Comp. 2 & 72.99 $\pm$ 0.03 & 73.03 $\pm$ 0.04 & 73.17 $\pm$ 0.02 & 71.32 $\pm$ 0.03 & 71.33 $\pm$ 0.01\\
& & Comp. 3 & 73.68 $\pm$ 0.01 & 73.67 $\pm$ 0.02 & 73.88 $\pm$ 0.01 & 74.37 $\pm$ 0.03 & 74.63 $\pm$ 0.01\\

\enddata
\tablecomments{Component column refers to the individual velocity components used to fit each profile. In all cases, Component 1 refers to the weak, broad component near $+63$\,km\,s$^{-1}$, while Components 2 and 3, respectively, refer to the lower and higher velocity components of the strong variable absorption around $+73$ km s$^{-1}$.}

\end{deluxetable*}
\capstarttrue

The normalized spectra were used in the line profile fitting program FITS6P \citep{welty94} to find the column density \emph{N},  line width parameter \emph{b}, and central velocity \emph{v} of each absorption component. Fits were made to both the static components between roughly \mbox{$-30$ km s$^{-1}$} and \mbox{$+30$ km s$^{-1}$} (Table 1), as well as the variable HV components at $v > 60$\,km\,s$^{-1}$ associated with the Monoceros Loop. Due to the high resolution (\mbox{$\Delta v \sim 1.3$\,km\,s$^{-1}$}) of this data, we account for the hyperfine splitting of the Na\,\textsc{i} D lines (\mbox{$\Delta v \sim 1.05$\,km\,s$^{-1}$}) when performing fits to these profiles. Fig. \ref{fig:full} shows the typical profile across the entire velocity range where absorption is observed. It is unclear if the myriad lower velocity components are also associated with the Monoceros Loop or some other intervening parcels of gas; however, no variation is seen in any of these components across our eight-year baseline, and thus they are excluded from our analysis.

For the high-velocity region, we fit each spectrum with three components - a weak, broad component at \mbox{$v \sim +63$\,km\,s$^{-1}$}, and the two variable components of interest at $v > +70$\,km\,s$^{-1}$. While the spectra of the first three epochs may appear by eye to be single components, we find that a fit composed of two components with slightly different central velocities yields a more accurate fit and smaller errors in all fit parameters, and is justified based on the clear two-component nature of the later epochs. 

\section{Results}
At high resolution, it becomes clear that the strong absorption near +70\,km\,s$^{-1}$ is actually two overlapping components. The values derived above provide clear evidence of dramatic variations in the Na\,\textsc{i} column density associated with this gas. These variations appear to occur in several stages (Fig. \ref{fig:hv}), beginning with an increase in \emph{N}(Na\,\textsc{i}) of $\sim$70\% (63\% and 80\% for the individual variable components at $+73$ and $+74$\,km\,s$^{-1}$, respectively) over a two-year period between 2006 and 2008, later followed by a $\sim$55\% decrease in \emph{N}(Na\,\textsc{i}) over a similar two-year period between 2009 and 2011. This decrease in column density is also accompanied by a velocity shift, with the velocity separation between the HV components increasing from $\sim$0.6 to $\sim$3.0\,km\,s$^{-1}$ over the same period. Finally, in the three-year period between 2011 and 2014, the column density again increased by $\sim$50\%, with the velocity separation further increasing to \mbox{$\sim$3.3\,km\,s$^{-1}$}.

\citet{welsh01} derive line parameters for the Na\,\textsc{i} components near \mbox{$+70$\,km\,s$^{-1}$} in their data, though their observations were taken at a velocity resolution that is larger that the velocity separation between the two components observed in our data (\mbox{\emph{R} $\sim$ 65,000}, corresponding to \mbox{$\Delta$v $\sim$ 4.5\,km\,s$^{-1}$}), and thus they treat this absorption as a single component. Nonetheless, by comparing their total column density of \mbox{$N=10.9\;(\pm1.1)\times 10^{11}$\,cm$^{-2}$} to the sum of the two components in our data (i.e. Components 2 and 3 in Table 2), we see that their data is most consistent with the largest value of column density in our sample, in 2009. Overall, our column densities tend to be less than the value reported in 2001, indicating that variation has been present in the $+73$\,km\,s$^{-1}$ component(s) since at least 2001.

The weak, broad component near $+63$\,km\,s$^{-1}$ also appears to exhibit weak variation across our observations. However, the exact size and shape of this component is susceptible to the placement of the continuum fit used during our data reductions. Therefore, differences in this continuum fit between spectra may manifest as apparent temporal variation across several spectra. Due to this effect, we are hesitant to assign much significance to the variations in this component. \citet{welsh01} derive values of \mbox{$N=0.5\;(\pm 6.6)\times 10^{11}$\,cm$^{-2}$} and \mbox{$b=1.1\;(\pm 3.9)$\,km\,s$^{-1}$}, which are consistent with the values presented in Table 2.

Despite the high-resolution data discussed here, the exact nature of this temporal variation remains unclear.
An analysis of LAB data \citep{kalberla05} and GALFA data \citep{galfa11} in the vicinity of HD~47240 reveals no apparent H\,\textsc{i} 21 cm emission at similar velocities to the varying HV gas, implying that the clouds responsible for this HV Na\,\textsc{i} absorption are likely much smaller than the GALFA beam size of $\sim$4\arcmin.
Additionally, of the 25 sightlines studied in this region by \citet{wallerstein76}, HD~47240 was the only one showing a component near \mbox{$+70$\,km\,s$^{-1}$}, indicating that this is not a pervasive component in the region.
\citet{lauroesch03} argued that Na\,\textsc{i}, as a trace neutral, would be biased to form on density peaks within the ISM since recombination rates exhibit a strong density dependence. Hence, it is possible that we are observing au-scale fluctuations in the physical conditions of the gas that cause \emph{N}(Na\,\textsc{i}) variations not necessarily representative of similar changes in \emph{N}(H\,\textsc{i}). Accounting for the \mbox{$\sim$1.3\,mas\,yr$^{-1}$} proper motion of HD~47240 and the \mbox{$\sim$50\,km\,s$^{-1}$} expansion velocity of the intervening gas, this sightline is sampling the gas within the Monoceros Loop on a spatial scale of $\sim$10\,au\,yr$^{-1}$.

Furthermore, the increasing velocity separation between the two components is indicative of gas dynamics beyond density fluctuations. These data support the conjecture by \citet{welsh01} that the high-velocity gas may be several shells with slightly different expansion velocities. Based on central velocities derived from our line fitting, the higher velocity component appears to accelerate by \mbox{$\sim$1\,km\,s$^{-1}$} over the course of our observations, while the lower velocity component decelerates by \mbox{$\sim$2\,km\,s$^{-1}$}. Of particular note is that this change did not occur gradually over the eight-year period of observation. The velocity of these components stayed essentially constant over three observations from 2006 to 2009, then changed very rapidly sometime during the approximately two-year period between our 2009 and 2011 observations, with a somewhat less dramatic change between 2011 and 2014. 

\section{Discussion}
The temporal variation of optical absorption has previously been reported toward the $\sim$10,000-year-old Vela SNR, initially by both \citet{hobbs91} toward HD~72127A and \citet{danks95} toward HD~72089 and HD~72997. This was later followed up more broadly by \citet{cha00}, who found evidence of Na\,\textsc{i} or Ca\,\textsc{ii} variability in 7 of 13 sightlines toward this region. \citet{kameswara15} then performed repeat observations toward 64 stars observed by \cite{cha00}, and found variation in the HV components for 4 of 11 sightlines in which high-velocity gas was observed. Due to the $>$100\,km\,s$^{-1}$ velocities of these components, they suggest a causal connection with the expanding SNR gas.
While most of the reported variations are in the equivalent widths of the absorption, several of these sightlines also exhibited variation in central velocity. Of particular note is the HD~72997 sightline, where \citet{danks95} found that the central velocity of the Na\,\textsc{i} $+190$ \mbox{km\,s$^{-1}$} component increased by \mbox{2.7\,km\,s$^{-1}$} over a 14-month period between 1991 and 1993, and \citet{cha00} noted a further \mbox{0.9\,km\,s$^{-1}$} between 1993 and 1996. These velocity changes are similar in size to those observed toward HD~47240, and occur over similar timescales.

These previous studies of the Vela SNR show that absorption lines that vary in their central velocity are not unheard of. However, HD~47240 is still curious due to its two overlapping high-velocity components, both of which begin at a similar velocity, but appear to change their central velocity in opposite directions over the same approximately two-year time frame. Two possible interpretations of this phenomenon are as follows. Either we are detecting a chance alignment between two unrelated parcels of gas that both undergo sudden velocity variations in the same time frame, or there is a single physical cause for this simultaneous variation, implying that these gas parcels are linked in some way. 

One possible explanation for the nature of these variations is that we are detecting cold gas forming as a result of interaction between colliding gas flows, similar to the results of \citet{vazquez06}. In their simulations of molecular cloud formation, they observed thin sheets of cold neutral material forming at the interface between warm diffuse flows. These cold thin sheets eventually became dynamically unstable and turbulent. Thus a sightline piercing such a region would simultaneously probe a cold, thin, turbulent sheet, as well as the warm, diffuse gas enveloping it. In all of our spectra of HD~47240, we detect a broad (i.e. warm) absorption component in addition to the narrow (i.e. cold) variable components. This broad component is much weaker than the variable components, as one would expect if this absorption was arising from diffuse, warm gas. If Na\,\textsc{i} is biased to form on density peaks within the ISM \citep{lauroesch03}, then we may be probing the interaction regions between the expanding SNR and the ambient ISM. 
Shocked gas in this interaction region could also explain the anomalous Na\,\textsc{i}/Ca\,\textsc{ii} ratios reported by \citet{wallerstein76}.

While this explanation, which suggests that the observed variations arise from a turbulent, cold sheet is feasible, we acknowledge that there are limitations to this interpretation. Simulations that try to capture the dynamics of a multiphase ISM \citep{vazquez06,vazquez11,gatto15} generally simulate large (10--100s of pc) regions with a limited resolution, typically no better than $\sim$0.03\,pc (corresponding to $\sim$6000\,au). \citet{vazquez06} specifically note that this resolution is "generally insufficient for correctly resolving the structure inside the cold layer during the initial stages of the evolution." Therefore, without a prediction for the typical size of gas structure in simulations, it is hard to say concretely if the observed fluctuations with a typical scale of 30\,au are consistent with the cold structures seen in simulations.

The rapid, dramatic Na\,\textsc{i} temporal variations observed toward the Monoceros Loop SNR make this region ideal for further optical and ultraviolet absorption line studies. Continued observations toward HD~47240, as well as additional observations of other sightlines in the direction of the SNR, will allow us to grasp the extent of the small-scale Na\,\textsc{i} structure in the high-velocity remnant gas. If these observations reveal that temporal variations such as those reported in this work are common in the Monoceros Loop, as in the Vela SNR, then this may suggest a connection between SNRs and small-scale Na\,\textsc{i} structures. Such a connection has previously been suggested by \citet{crawford03}, and observations by \citet{meyer15} found Na\,\textsc{i} variations in 12 of 20 sightlines, with almost all of these variable sightlines being associated with known H\,\textsc{i} shells, SNRs, or stellar bow shocks. A broad survey by \citet{mcevoy15} of 104 sightlines with existing spectra from VLT-UVES and the McDonald Observatory found only $\sim$6\% of sightlines to exhibit temporal variation, primarily in the Na\,\textsc{i} D lines. These results suggest that this effect may not be pervasive, but rather localized to shell-like ISM structures. However, more investigation is currently needed to understand the prevalence of au-scale Na\,\textsc{i} fluctuations in the diffuse ISM, as well as to understand the underlying mechanisms that give rise to these structures.

\section*{Acknowledgements}
We wish to thank Daryl Willmarth for his assistance with the Kitt Peak National Observatory observations. This research has made use of the SIMBAD database, operated at CDS, Strasbourg, France. We thank the referee for helpful comments that improved this paper.

\emph{Facilities:} \facility{KPNO:CFT}

\clearpage

\end{document}